The world of crypto-currency is currently driven by Bitcoin due to its support from ASICs. There is an opportunity to build ASICs for Scrypt Cryptocurrencies so that they can flourish as well.

# Scrypt Mining with ASICs

By David Watkins

Under supervision of Simha Sethumadhavan

# Introduction

Crypto-currencies have garnered a lot of attention by governments and internet enthusiasts over the past three years. These currencies are celebrated for their security and speedy transactions in a modern era of digital commerce. Bitcoin was the first of these currencies to gain a large advantage over subsequent iterations. Bitcoin was first conceived by Satoshi Nakamoto who mentioned the concept of a crypto-currency in his paper titled Bitcoin. It featured new concepts such as proof of work and transactions which utilized hash-based encryption.

One particular alternative crypto-currency is known as Litecoin. Backed by a memory-intensive algorithm known as Scrypt, many crypto-currency enthusiasts have decided to celebrate this particular coin. Scrypt expands on Bitcoin's proof of work algorithm by adding the amount of work it takes to commit a transaction within the Litecoin network. Scrypt forces more work on the device that is being used to perform the algorithm by making frequent memory requests. This makes it difficult to create specialized hardware to create new coins and to commit transactions due to the nature of memory intensive applications.

Creating specialized hardware is very important for the overall health of a crypto-currency. It allows for investors and third parties to help grow the currency into a valuable form of exchange. Because Bitcoin's proof of work algorithm is not as intensive, it was much easier to create an ASIC (Application Specific Integrated Circuit) to help mine those coins faster. Third parties were able to invest in Bitcoin's growth and it is now the most valuable crypto-currency in existence today. Due to the nature of Litecoin, however, the ASIC development has not been as fast. Many people have requests ASICs be made for Scrypt, but there has not been any development until now.

As part of this research project, different solutions and strategies for developing a Litecoin ASIC were explored. The research began with finding and compiling Litecoin source code. Determining which portion of the Scrypt algorithm was the most intensive was important for determining how best to optimize the circuit. Compiling the source code using different optimization strategies was useful for determining what portion of the code would be able to be sped up or perhaps designated a special portion of the circuit.

The most expensive part of the algorithm is the memory-hard salsa20. In order to reduce the overall impact that this portion of the Scrypt algorithm had on the overall runtime, different memory solutions to be implemented into a cache were explored. By utilizing the proximity and speed of a cache for this particular circuit, the overall speed of the algorithm would be drastically increased. Different memory technologies, such as SRAM, STT-RAM, MRAM, PCRAM, and even DRAM were explored as potential candidates. Due to the experimental nature of some of these options, SRAM was found to be the fastest candidate.

The final portion of the research was to design and benchmark a schematic of the ASIC in Verilog code. There were some significant benefits to producing the ASIC, including reduced cost for production over current alternatives. The cost of production and potential next steps were explored in order to generate an idea of what it would take to actually produce a real ASIC. It was important to take into consideration the time to create an ASIC given the rapidly changing market of crypto-currencies as well and the upcoming competition in the ASIC market.

# Background

Crypto-currencies are a cross between a fiat currency and a commodity. They use cryptography to ensure safe and secure transaction between parties[1]. Crypto-currencies have garnered much attention over the past four years due to their revolutionary method of how currency can be exchanged and used digitally, bypassing centralized government and financial institutions. The inception of these currencies began with a paper written by Satoshi Nakamoto who wrote "Bitcoin: A Peer-to-Peer Electronic Cash System". He suggested a new system that allowed online payments to be sent directly from one party to another without a need for a financial institution (Nakamoto, 1).

## Satoshi Nakamoto's Peer to Peer Coin

By replacing online electronic transactions with crypto-currencies, people would not have to be as concerned about their online information being publicly available. Each transaction required chain of digital signatures by way of public and private keys. The public key could be well known as a destination for coins, but the private key is necessary to be a source of coins. Each transaction would be protected by a proof of work, or hashing algorithm that would prove that the transaction is valid. The only way a user could submit their exchange would be by performing the hashing algorithm on their end of the computer. Each crypto-currency was to have its own blockchain, a public ledger, of all transactions that have occurred on the network so far. Fraud cannot occur in such a system.

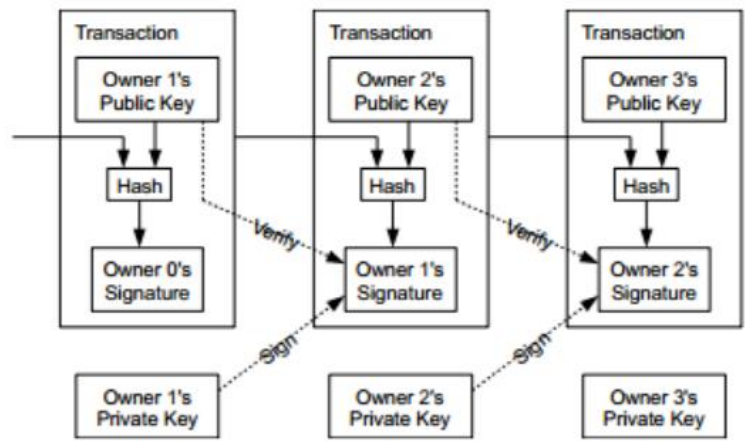

Figure 1 - Each Owner's public key and private key are used to provide a signature of a transaction to owners that appear later in the chain of transactions (Nakamoto, 2)

The hashing algorithm that is primarily used by Bitcoin is known as SHA-256. This algorithm, in 2009, was originally quite secure and had not been easily cracked. SHA-256 is a very popular hashing algorithm for many crypto-currencies. Collisions in the Bitcoin block chain would cause an overlap of transaction data. But because of hashing, the odds of having a collision in the blockchain are minimal. The odds of a collision happening in SHA-256 are $\frac{1}{2^{128}}$. These odds, multiplied by the approximate 52500 transactions that happen per year for Bitcoin, are minimal enough that a collision will likely never occur.

---

[1] http://www.amazon.com/Bitcoin-Revolution-Ending-Tyranny-Profit-ebook/dp/B00CAX5OZQ

Bitcoin's popularity as a currency has been due to many exchanges and users that have grown to support and mine in the network. The open source nature of the coin allowed users to design versatile program such as CPUMiner and CGMiner which could both be used to mine these coins. Various alternate currencies were created in response to this excitement. These included Namecoin and Litecoin. Litecoin in particular was created with a different proof of work algorithm known as Scrypt.

## Scrypt and How it Improves Bitcoin's Proof of Work

Litecoin was the first Crypto-currency to popularize the idea of Scrypt as an alternative to SHA-256 based coins. Bitcoin ASICs made it difficult for new, Sha-256 based Alt Coins to gain popularity because their difficulty was going up so fast that large communities of people were not able to get large quantities of the currency via solo mining. Scrypt Cryptocurrencies, however, do not have large support via ASICs yet. Scrypt Alt Coins are typically mined solely by GPU miners. These alternative Cryptocurrencies have a market cap much smaller than that of Bitcoin[2]. Many attribute this lack of value to the lack of investment in Scrypt coins and lack of ASICs. The current lack of ASICs for Scrypt Cryptocurrencies are due to the nature of Scrypt and how it creates a larger computational problem than solely Sha-256.

## What is Scrypt?

Scrypt was suggested by Colin Percival as a memory-hard algorithm to help create a system for password-key derivation. In his paper from May 2009, he describes a new methodology for hashing values that is not easily brute-forced. Key derivation functions are functions that will generate a derived key from a password. This key can be generated dynamically every time a password is entered, and tested with the original derived key to determine if the keys are a match. This is superior to storing a single password because it does not expose the actual password to potential thieves (Percival, 2).

In order to break key derivation functions, a hacker might attempt to find user passwords by hacking into a server and pulling all database info about all users. For most key derivation functions, the only feasible option is to test all possible passwords until the correct value is found, otherwise known as a "brute force" attack[3]. The rate at which a hacker can test passwords is linearly related to the length of time it takes to derive a key to test if the password is correct. This way, if the designer of the algorithm can make the algorithm twice as long, it will take twice as long to crack the password. Originally, Sha-256 was a very good candidate for such functionality, but soon after it was popularized many computer architects were able to determine how to make ASICs to solve SHA-256 passwords quickly.

In order to combat this potential flaw in key derivation functions, the concept of "memory-hard" algorithms was invented. A memory hard algorithm will need to access random-access-memory, or RAM, frequently as part of its runtime. Typically, these queries can add significant amount of time to the overall runtime of a circuit. A typical access time to DRAM (Dynamic RAM) will take 60ns[4]. If a key derivation function needs to make 1000 accesses to DRAM per password, this could mean an additional 60μs per password. This time is small for one particular password, but a brute-force attack that tests for combinations of English words would need an additional 3.9 days[5] to compute. This estimate does not include time spent on the algorithm, just purely from a memory access perspective.

---

[2] As of this writing, the coin market cap of Bitcoin is $5,654,660,751 versus Litecoin which is only $296,745,377. These two currencies have the highest market caps out of all other Cryptocurrencies.
[3] http://searchsecurity.techtarget.com/definition/brute-force-cracking

There are multiple memory-hard algorithms that could be used for Scrypt, but Percival recommended the use of ROMix in particular because it is a sequentially, memory-hard algorithm (Percival, 7). ROMix acts as if it is randomly accessing memory and does so in a sequential way. The algorithm for ROMix is difficult to parallelize. This is good because a sequential algorithm will take longer to perform than one that can have all its individual parts performed simultaneously.

## Why Litecoin Creators Chose Scrypt

Scrypt is a memory-hard function. Colin Percival argued heavily that creating an ASIC for Scrypt would be expensive and unrealistic. The Scrypt algorithm works as a proof-of-work scheme similar to how Sha-256 worked for Bitcoin. The main difference is that Scrypt prevents ASICs and GPU mining, which were very effective for mining Bitcoin, from having any efficient power over Scrypt. In April 2014, Litecoin reached a market cap of $270 million and achieved a network strength of 100GH/s. This equates to 100 TH/s for Bitcoin due to the intensity that Scrypt imposes on the hardware mining it[6].

Because Scrypt is inherently more difficult to perform computationally, many Alt Coins have adopted it as the proof-of-work scheme. Dogecoin, a coin popularized by many social networking sites, has been one such Crypto-currency that has gained a lot of momentum in only six months[7]. Many people argue that these coins would not be popular if Scrypt were easily mined. This view does not take into account that ASICs are what helped Bitcoin rise to its current popularity and allowed it to gain a much higher market cap value. Without ASICs, these coins will not flourish to their full potential.

# Developing a Scrypt ASIC

The first steps taken towards the development of the Scrypt ASIC were to find and compile the mining program responsible for performing Scrypt. This program was then benchmarked and the time spent in each portion of the program was calculated. It was quickly apparent that this particular algorithm was memory-hard. Looking for good alternatives to traditional hardware was important. In particular, exploring a cache dedicated to storing the Litecoin block header was key. Different memory alternatives were then analyzed to see which would be the most appropriate for this particular application. Finally, a sample circuit designs was created using Verilog and then benchmarked further to see how expensive the design would be.

---

[4] http://www.techopedia.com/definition/2770/dynamic-random-access-memory-dram
[5] There are about 75,000 words, try two word combinations: $75,000^2$. Multiply by $60*10^{-6}$. Then divide those seconds into days, and the result is 3.90625 days.
[6] https://en.bitcoin.it/wiki/Litecoin
[7] http://dogecoin.com/

# Scrypt in Theory

Developing a Scrypt ASIC is not an easy task, however. As Percival noted in his paper, the algorithm is designed to be memory-intensive. However, since the time of his writing, the cost to implement certain circuits has gone down enough that developing a Litecoin ASIC may prove to be cost effective. In particular, the cost of developing the memory can be reduced once the designer is aware that there only needs to be 128KB of memory per processor. Because Scrypt is not easily parallelizable, it is not necessary to have multiple cores, but instead have one core driving a 128KB cache.

# Scrypt Call-Flow

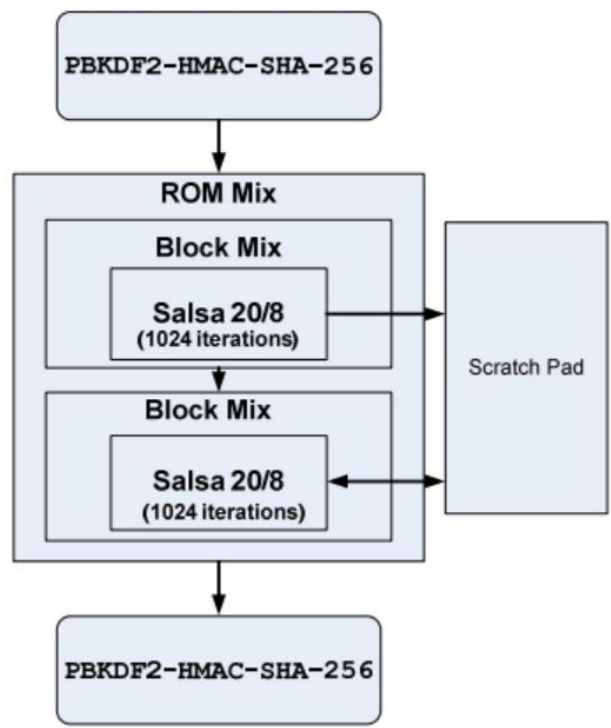

Figure 2 Scrypt algorithm call flow. Begins with a Sha-256 hash. Then performs a ROM Mix, Block Mix, and 1024 iterations of Salsa20/8, twice. Then Sha-256 hashes the result again. (Alpha-Technologies, 2)

Scrypt begins with a basic key derivation function, Sha-256. This value is 128 kB long. This is the only amount of space that the algorithm needs to complete. Each Salsa iteration requires a total of 34 cycles. One cycle to read in 128 bytes of the overall hashed value. The inside loop (see Appendix A) has two halves, each with four sections that can be done in parallel. This means one iteration of inner loop will take 8 cycles, and each loop has 4 iterations, so a total of 32 extra cycles. And then an additional cycle for the final loop.

In both iterations, there are 2048 calls to salsa because each loop makes two calls. The second iteration also has to access the cache to get the new values so that it can continue. Theoretically, this overall process could take 139264 cycles to complete one iteration of Scrypt plus whatever latency would be required for accessing memory. If one cycle in this circuit took 1 ns, that could mean a potential hash rate of 7 KH/s. This is ignoring potential speedup from additional parallelization and potential extra speedup from a faster clock cycle.

## Scrypt on a CPU

Scrypt runs very differently depending on the hardware that is performing the computation. The trials performed primarily ran the program on a processor. The program was run on a computer that had Ubuntu installed and was also compiled for the computer[8]. The code was also compiled using a variety of different optimization strategies. In particular: msse2, msse4, O4, and O3 compiler flags were used to analyze the particular performance improvements that could be observed on standard commercial hardware. By far the fastest optimization strategy was MSSE4 which had a 35% improvement on total runtime over no compiler optimization[9].

Due to the limitation of the testing environment, there was no way to run the program without the OS's (in this case Ubuntu's) normal programs taking up CPU clock cycles. Increasing performance using MSSE4 meant that the program was being optimized for use of all four available cores on the computer[10]. The memory hard portion of the program, in this case the Salsa20 Core function, was taking the most processing time. The code is as follows:

```c
for (i = 0; i < 8; i += 2) {
    /* Operate on columns. */
    x04 ^= ROTL(x00 + x12,  7);   x09 ^= ROTL(x05 + x01,  7);
    x14 ^= ROTL(x10 + x06,  7);   x03 ^= ROTL(x15 + x11,  7);

    x08 ^= ROTL(x04 + x00,  9);   x13 ^= ROTL(x09 + x05,  9);
    x02 ^= ROTL(x14 + x10,  9);   x07 ^= ROTL(x03 + x15,  9);

    x12 ^= ROTL(x08 + x04, 13);   x01 ^= ROTL(x13 + x09, 13);
    x06 ^= ROTL(x02 + x14, 13);   x11 ^= ROTL(x07 + x03, 13);

    x00 ^= ROTL(x12 + x08, 18);   x05 ^= ROTL(x01 + x13, 18);
    x10 ^= ROTL(x06 + x02, 18);   x15 ^= ROTL(x11 + x07, 18);

    /* Operate on rows. */
    x01 ^= ROTL(x00 + x03,  7);   x06 ^= ROTL(x05 + x04,  7);
    x11 ^= ROTL(x10 + x09,  7);   x12 ^= ROTL(x15 + x14,  7);

    x02 ^= ROTL(x01 + x00,  9);   x07 ^= ROTL(x06 + x05,  9);
    x08 ^= ROTL(x11 + x10,  9);   x13 ^= ROTL(x12 + x15,  9);

    x03 ^= ROTL(x02 + x01, 13);   x04 ^= ROTL(x07 + x06, 13);
    x09 ^= ROTL(x08 + x11, 13);   x14 ^= ROTL(x13 + x12, 13);

    x00 ^= ROTL(x03 + x02, 18);   x05 ^= ROTL(x04 + x07, 18);
    x10 ^= ROTL(x09 + x08, 18);   x15 ^= ROTL(x14 + x13, 18);
}
```

---

[8]Code was obtained from: https://github.com/pooler/cpuminer
[9] Runtime was 2.56ms for no optimization versus 1.89 ms for msse4 optimization.
[10] Intel Core i7 2630QM

The code uses variable placeholder for values in memory so that the code can be easily parallelized. Each section of four functions can be run in parallel together as there are no interdependencies on the code. If the proper parallelization flags are set during compile time, the program has at least a four time speedup. This portion of the algorithm takes approximately 60% of the runtime for the entire Scrypt runtime.

38% of the runtime is devoted to memory access. The Salsa20 algorithm begins with 32 reads and 16 writes from memory where it reads 16 memory locations of the two parameters passed to the function:

```
x00 = (B[ 0] ^= Bx[ 0]);
x01 = (B[ 1] ^= Bx[ 1]);
x02 = (B[ 2] ^= Bx[ 2]);
x03 = (B[ 3] ^= Bx[ 3]);
x04 = (B[ 4] ^= Bx[ 4]);
x05 = (B[ 5] ^= Bx[ 5]);
x06 = (B[ 6] ^= Bx[ 6]);
x07 = (B[ 7] ^= Bx[ 7]);
x08 = (B[ 8] ^= Bx[ 8]);
x09 = (B[ 9] ^= Bx[ 9]);
x10 = (B[10] ^= Bx[10]);
x11 = (B[11] ^= Bx[11]);
x12 = (B[12] ^= Bx[12]);
x13 = (B[13] ^= Bx[13]);
x14 = (B[14] ^= Bx[14]);
x15 = (B[15] ^= Bx[15]);
```

Each memory location is read and xor'd with the corresponding location in the second parameter of the function. These values are then written back to the B array and also stored locally in variables defined at runtime. A similar write-back sequence is performed at the end of the salsa 20 function:

```
B[ 0] += x00;
B[ 1] += x01;
B[ 2] += x02;
B[ 3] += x03;
B[ 4] += x04;
B[ 5] += x05;
B[ 6] += x06;
B[ 7] += x07;
B[ 8] += x08;
B[ 9] += x09;
B[10] += x10;
B[11] += x11;
B[12] += x12;
B[13] += x13;
B[14] += x14;
B[15] += x15;
```

The final resulting values that were computed in the central algorithmic portion of the salsa 20 function are then stored into the corresponding location in the B array. This features another 16 reads and 16 writes, as each value is incremented by the values stored locally. Altogether there are 48 reads and 32 writes for the salsa20 function. If this time that it takes for the computer to write-back this data can be reduced, it will speed up the program significantly.

## Memory Considerations

Deciding on which memory technology to use to implement the Scrypt ASIC is important. There are multiple flavors of memory that could be used. The most important piece to note is that cache memory is much faster than RAM. The typical DRAM speed is approximately 60 ns, whereas SRAM can achieve access times of 0.75 ns (See Appendix B). After analyzing the Scrypt function, the data showed where the majority of the time was spent computationally[11].

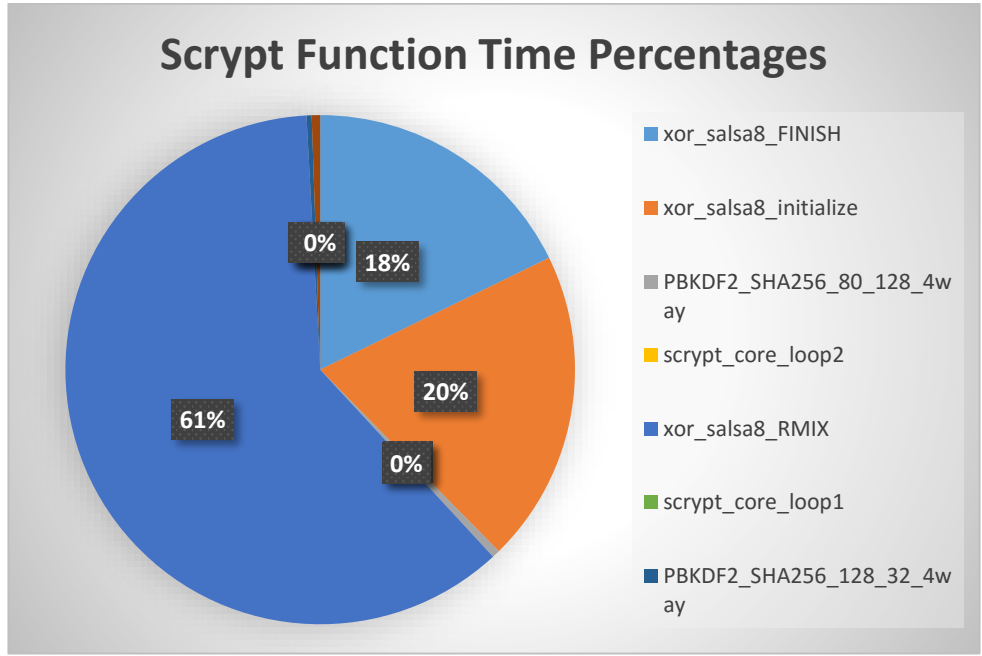

**Figure 3** This data shows where the majority of the time spent in Scrypt was. In particular, the ROMix portion of Salsa too the most time.

In order to reduce the overall time for the ASIC to perform the task of mining, it is necessary to analyze the most common case for the algorithm and minimize the time spent. In particular, optimizing the program with compiler flags such as msse4 and O4 both helped optimize the speed and decreased the amount of time the program took by approximately 1 second (See Appendix D). There are many memory structures that could potentially increase the overall speed of the ASIC, such as STT-RAM, PC-RAM, and MRAM.

## SRAM

Static random access memory (SRAM) is a type of semiconductor memory that uses a latch based circuit to store information in the form of bits. SRAM is a volatile storage, or it does not maintain the data across cycling of power for a given memory cell. SRAM is not very power intensive, as it only needs a few microwatts when running to maintain power. SRAM is a good candidate for cache because it maintains value when not accessed for relatively long periods of time and its very quick. After analyzing output from Cacti 6.0, the data showed that the read speed of 0.75 ns. This is very quick and can allow for a maximum clock cycle of 1.33GHz. (Dillibabu)

---

[11] The CPU miner program was run on a computer that was connected to a Litecoin testnet. This Litecoin testnet provided a sample block header with relatively low difficulty (<1). The program then reported the number of hashes as well as how long it spent in each function. Overall the program ran 4104 hash for one block.

## MRAM

MRAM is a memory technology that has existed since the 1990s and has a higher density from flash RAM and DRAM. MRAM relies on magnetic fields to maintain the value held by a given memory cell. It is non-volatile, so it lasts for long periods of time without losing data. MRAM has inferior read and write speeds to SRAM, which makes it a poor candidate for this ASIC. (Tamai)

## STT-RAM

Spin-transfer torque RAM (STT-RAM) is a form of RAM that combines the speed of SRAM with the permanence of flash memory. It relies a magnetic tunnel junction, or a way to transmit magnetic fields, to modify a state using a spin-polarized current. In other words, the value in the STT-RAM is held constant until it receives a magnetic signal changing the state. This can be useful for maintaining values held in memory for long periods of time. However, this technology is not quite as fast as SRAM. STT-RAM has a write latency of 10.61ns, twenty times longer than that of SRAM. This would not make a good candidate for speeding up the runtime of the ASIC as it would be much more expensive and slower than using pre-existing technology. (Penn State)

## PC-RAM

A type of non-volatile random-access memory. Similar to STT-RAM, it does not expire quickly. It relies on a complex 3D lattice system to produce long lasting data. However, its latency varies between 10-100ns for both reading and writing data. Due to its inconsistency and lack of real implementation, it would be a poor choice for implementation. (Xie)

## Preliminary Designs and Estimates

After performing minor modifications to an online source code[12] for an FPGA Scrypt miner, I was able to determine the number of logical gates and time delay of the overall circuit. There are 13,446 logical gates on the chip, where 12,613 of them are combinational function and 6,567 are logic registers. These logical registers have a total of 1,048,576 memory bits to store for one operation. This is 131,072 bytes, or roughly a 128KB cache.

To analyze the timing of the circuit, I found the worst case timing path for one operation to find the worst case operation time. In this case the worst case path was 95.762 ns. This timing could allow for at least one hash to occur, resulting in an overall timing of 10,526,315 H/s. This speed is much higher than the best graphics cards currently, which only clock in at 1 MH/s[13].

Given the sheer number of logical circuits, however, the overall cost of this ASIC would be very high. According to Multi-Project Circuits, the cost per mm² of producing an ASIC would be approximately 3500 Euro for a 130nm chip. Given that there are 6,567 SRAM cache registers which each would be 5.007 $\mu$m²/bit[14], this would result in approximately 32883.6 $\mu$m² for the cache alone. The combinational circuit would likely feature an average size of $5\mu m^2$, for a total approximate circuit area of $227483.6\mu m^2$ or $0.22748mm^2$. This chip would initially be about 796.18 Euro to produce, or 1087.26 USD.

---

[12] https://github.com/kramble/FPGA-Litecoin-Miner
[13] https://litecoin.info/Mining_hardware_comparison
[14] http://www.cl.cam.ac.uk/research/srg/han/ACS-P35/obj-4.2/zhpd802be4b0.html

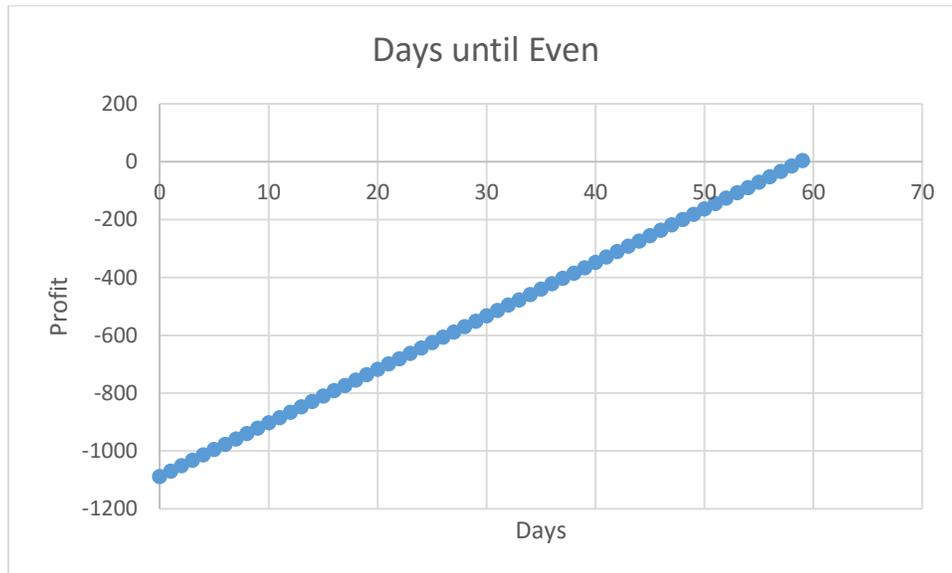

**Figure 4 it will take approximately 60 days to break even assuming no fluctuations in the most profitable coin's value**

This chip would be able to produce 10 MH/s. At the currently most profitable Scrypt coins[15], this would result in $18.50 of revenue per day, not including power costs. This would take $60 to break even at current valuations of the most profitable coins. If their value were to increase, the number of days would fall as a result. These costs do not take into account the amount of money it would take to pay a team of people to produce the ASIC. This only takes into account potential fabrication facilities' rates of producing one ASIC. These costs will likely go down as the cost of producing 130nm chips declines over the coming years.

## How a Scrypt ASIC will better the Alt Coin Economy

By providing an ASIC for Scrypt Cryptocurrencies, there will be a resurgence of following in these Alt Coins that will be akin to that of Bitcoin. People will purchase the ASICs which will help produce more ASICs and also allow for additional investment in the Crypto-currency itself. By offering faster coin production and higher investment rates, the overall market cap will increase, thus increasing the probability that the coin will expand in intrinsic value. It is possible, that with the number of Scrypt Alt Coins in existence, that the overall market cap for all Scrypt coins will be greater than that of Bitcoin.

Many new currencies have come about recently to combat the incoming ASICs in the Scrypt market. These new crypto-currencies rely on new proof-of-work algorithms such as Scrypt-N[16] and Quark[17]. Scrypt-N relies on modulating the number of SHA-256 hashes which would be difficult to implement using an ASIC. Quark relies on multiple hashing algorithms per round, preventing easy fabrication of the same circuit. These combatants do not realize the support that ASICs provide for the crypto-currency market. By allowing entrepreneurs to invest money in ASICs, the market can grow in value. Continuing to complicate and saturate the market with new coins and new proof-of-work algorithms will only proceed to fragment and potentially cripple the community.

---

[15] http://www.coinwarz.com/cryptocurrency/?sha256hr=1.00&sha256p=100.00&sha256pc=0.1000&scrypthr=10000.00&scryptp=500.00&scryptpc=0.1000&scryptnhr=450.00&scryptnp=500.00&scryptnpc=0.1000&x11hr=3000.00&x11p=500.00&x11pc=0.1000&keccakhr=500.00&keccakp=500.00&keccakpc=0.1000&quarkhr=5000.00&quarkp=500.00&quarkpc=0.1000&groestlhr=10.00&groestlp=500.00&groestlpc=0.1000&sha256c=false&scryptc=true&scryptnc=false&x11c=false&keccakc=false&quarkc=false&groestlc=false&e=Coinbase

[16] http://give-me-coins.com/support/n-scrypt-guide/

[17] http://www.qrk.cc/

## CPU Mining and GPU Mining Fall Short

CPU Mining, as noted earlier, is very slow compared to many alternatives. The runtime for the earlier program was roughly 4.1KH/s and a more modern processor would be able to produce hashes at 95KH/s. The same program running on a much more powerful GPU, such as an AMD R9 290X would show speeds of approximately 804 KH/s[18]. These speeds pale in comparison to the theoretical 10MH/s that the ASIC design is capable of providing. This is roughly a 1000 times increase over CPU and a 10 times increase over GPU miners, keeping into consideration possible fluctuations in hashing rates.

Not only is the effective hashrate of consideration, but the cost of energy for each of these designs is important as well. A CPU and GPU needs to be housed within a computer enclosure that is capable of providing energy and instructions to both in order to run the program. The Core-i7-4770 runs at 84W and the AMD R9 290X runs at approximately 300W. These would incur additional costs above their entry cost in the form of electricity bills. This particular ASIC will only use 300W to perform its computation[19]. The power advantage over these alternatives is apparent, given that it would take ~84,000W for a cluster of CPUs to reach comparable performance and 3000W for a cluster of GPUs to reach similar performance to the ASIC.

# Conclusion

Research in the development of ASICs is a time consuming process. Steps taken towards this goal included compiling and benchmarking Scrypt code, analyzing various components involved with the ASIC including memory, and analyzing costs associated with development of the ASIC. Litecoin ASIC development will also require further research in the best fabrication facilities and development of a functioning prototype. This prototype would likely be made using an FPGA circuit.

The next steps in developing the ASIC would be to first start pitching the idea to VCs to not only raise awareness about the issue but also to raise funds to get a team together to start designing the ASIC. Ultimately the current design will likely be redone so that it is more efficient and relies on the experience of a team that has built an ASIC before. As soon as these steps are complete, determining the best foundry to fabricate the ASIC or potentially get a prototype would be key to making sure that the ASIC is fully functional.

---

[18] Speeds for various mining hardware obtained from: https://litecoin.info/Mining_hardware_comparison
[19] Estimate based on current power consumption of comparable ASIC designs: https://en.bitcoin.it/wiki/Mining_hardware_comparison

# Appendix A – Scrypt Algorithm in C++

Code obtained from: https://github.com/litecoin-project/litecoin

```
/*
 * Copyright 2009 Colin Percival, 2011 ArtForz, 2012-2013 pooler
 * All rights reserved.
 *
 * Redistribution and use in source and binary forms, with or without
 * modification, are permitted provided that the following conditions
 * are met:
 * 1. Redistributions of source code must retain the above copyright
 *    notice, this list of conditions and the following disclaimer.
 * 2. Redistributions in binary form must reproduce the above copyright
 *    notice, this list of conditions and the following disclaimer in the
 *    documentation and/or other materials provided with the distribution.
 *
 * THIS SOFTWARE IS PROVIDED BY THE AUTHOR AND CONTRIBUTORS ``AS IS'' AND
 * ANY EXPRESS OR IMPLIED WARRANTIES, INCLUDING, BUT NOT LIMITED TO, THE
 * IMPLIED WARRANTIES OF MERCHANTABILITY AND FITNESS FOR A PARTICULAR PURPOSE
 * ARE DISCLAIMED.  IN NO EVENT SHALL THE AUTHOR OR CONTRIBUTORS BE LIABLE
 * FOR ANY DIRECT, INDIRECT, INCIDENTAL, SPECIAL, EXEMPLARY, OR CONSEQUENTIAL
 * DAMAGES (INCLUDING, BUT NOT LIMITED TO, PROCUREMENT OF SUBSTITUTE GOODS
 * OR SERVICES; LOSS OF USE, DATA, OR PROFITS; OR BUSINESS INTERRUPTION)
 * HOWEVER CAUSED AND ON ANY THEORY OF LIABILITY, WHETHER IN CONTRACT, STRICT
 * LIABILITY, OR TORT (INCLUDING NEGLIGENCE OR OTHERWISE) ARISING IN ANY WAY
 * OUT OF THE USE OF THIS SOFTWARE, EVEN IF ADVISED OF THE POSSIBILITY OF
 * SUCH DAMAGE.
 *
 * This file was originally written by Colin Percival as part of the Tarsnap
 * online backup system.
 */

#include "scrypt.h"
#include "util.h"
#include <stdlib.h>
#include <stdint.h>
#include <string.h>
#include <openssl/sha.h>

#if defined(USE_SSE2) && !defined(USE_SSE2_ALWAYS)
#ifdef _MSC_VER
// MSVC 64bit is unable to use inline asm
#include <intrin.h>
#else
// GCC Linux or i686-w64-mingw32
#include <cpuid.h>
#endif
#endif

static inline uint32_t be32dec(const void *pp)
{
        const uint8_t *p = (uint8_t const *)pp;
        return ((uint32_t)(p[3]) + ((uint32_t)(p[2]) << 8) +
            ((uint32_t)(p[1]) << 16) + ((uint32_t)(p[0]) << 24));
}
```

```c
static inline void be32enc(void *pp, uint32_t x)
{
        uint8_t *p = (uint8_t *)pp;
        p[3] = x & 0xff;
        p[2] = (x >> 8) & 0xff;
        p[1] = (x >> 16) & 0xff;
        p[0] = (x >> 24) & 0xff;
}

typedef struct HMAC_SHA256Context {
        SHA256_CTX ictx;
        SHA256_CTX octx;
} HMAC_SHA256_CTX;

/* Initialize an HMAC-SHA256 operation with the given key. */
static void
HMAC_SHA256_Init(HMAC_SHA256_CTX *ctx, const void *_K, size_t Klen)
{
        unsigned char pad[64];
        unsigned char khash[32];
        const unsigned char *K = (const unsigned char *)_K;
        size_t i;

        /* If Klen > 64, the key is really SHA256(K). */
        if (Klen > 64) {
                SHA256_Init(&ctx->ictx);
                SHA256_Update(&ctx->ictx, K, Klen);
                SHA256_Final(khash, &ctx->ictx);
                K = khash;
                Klen = 32;
        }

        /* Inner SHA256 operation is SHA256(K xor [block of 0x36] || data). */
        SHA256_Init(&ctx->ictx);
        memset(pad, 0x36, 64);
        for (i = 0; i < Klen; i++)
                pad[i] ^= K[i];
        SHA256_Update(&ctx->ictx, pad, 64);

        /* Outer SHA256 operation is SHA256(K xor [block of 0x5c] || hash). */
        SHA256_Init(&ctx->octx);
        memset(pad, 0x5c, 64);
        for (i = 0; i < Klen; i++)
                pad[i] ^= K[i];
        SHA256_Update(&ctx->octx, pad, 64);

        /* Clean the stack. */
        memset(khash, 0, 32);
}

/* Add bytes to the HMAC-SHA256 operation. */
static void
HMAC_SHA256_Update(HMAC_SHA256_CTX *ctx, const void *in, size_t len)
{
        /* Feed data to the inner SHA256 operation. */
        SHA256_Update(&ctx->ictx, in, len);
```

```c
}

/* Finish an HMAC-SHA256 operation. */
static void
HMAC_SHA256_Final(unsigned char digest[32], HMAC_SHA256_CTX *ctx)
{
	unsigned char ihash[32];

	/* Finish the inner SHA256 operation. */
	SHA256_Final(ihash, &ctx->ictx);

	/* Feed the inner hash to the outer SHA256 operation. */
	SHA256_Update(&ctx->octx, ihash, 32);

	/* Finish the outer SHA256 operation. */
	SHA256_Final(digest, &ctx->octx);

	/* Clean the stack. */
	memset(ihash, 0, 32);
}

/**
 * PBKDF2_SHA256(passwd, passwdlen, salt, saltlen, c, buf, dkLen):
 * Compute PBKDF2(passwd, salt, c, dkLen) using HMAC-SHA256 as the PRF, and
 * write the output to buf.  The value dkLen must be at most 32 * (2^32 - 1).
 */
void
PBKDF2_SHA256(const uint8_t *passwd, size_t passwdlen, const uint8_t *salt,
    size_t saltlen, uint64_t c, uint8_t *buf, size_t dkLen)
{
	HMAC_SHA256_CTX PShctx, hctx;
	size_t i;
	uint8_t ivec[4];
	uint8_t U[32];
	uint8_t T[32];
	uint64_t j;
	int k;
	size_t clen;

	/* Compute HMAC state after processing P and S. */
	HMAC_SHA256_Init(&PShctx, passwd, passwdlen);
	HMAC_SHA256_Update(&PShctx, salt, saltlen);

	/* Iterate through the blocks. */
	for (i = 0; i * 32 < dkLen; i++) {
		/* Generate INT(i + 1). */
		be32enc(ivec, (uint32_t)(i + 1));

		/* Compute U_1 = PRF(P, S || INT(i)). */
		memcpy(&hctx, &PShctx, sizeof(HMAC_SHA256_CTX));
		HMAC_SHA256_Update(&hctx, ivec, 4);
		HMAC_SHA256_Final(U, &hctx);

		/* T_i = U_1 ... */
		memcpy(T, U, 32);

		for (j = 2; j <= c; j++) {
```

```c
                /* Compute U_j. */
                HMAC_SHA256_Init(&hctx, passwd, passwdlen);
                HMAC_SHA256_Update(&hctx, U, 32);
                HMAC_SHA256_Final(U, &hctx);

                /* ... xor U_j ... */
                for (k = 0; k < 32; k++)
                    T[k] ^= U[k];
            }

            /* Copy as many bytes as necessary into buf. */
            clen = dkLen - i * 32;
            if (clen > 32)
                clen = 32;
            memcpy(&buf[i * 32], T, clen);
        }

        /* Clean PShctx, since we never called _Final on it. */
        memset(&PShctx, 0, sizeof(HMAC_SHA256_CTX));
}

#define ROTL(a, b) (((a) << (b)) | ((a) >> (32 - (b))))

static inline void xor_salsa8(uint32_t B[16], const uint32_t Bx[16])
{
        uint32_t x00,x01,x02,x03,x04,x05,x06,x07,x08,x09,x10,x11,x12,x13,x14,x15;
        int i;

        x00 = (B[ 0] ^= Bx[ 0]);
        x01 = (B[ 1] ^= Bx[ 1]);
        x02 = (B[ 2] ^= Bx[ 2]);
        x03 = (B[ 3] ^= Bx[ 3]);
        x04 = (B[ 4] ^= Bx[ 4]);
        x05 = (B[ 5] ^= Bx[ 5]);
        x06 = (B[ 6] ^= Bx[ 6]);
        x07 = (B[ 7] ^= Bx[ 7]);
        x08 = (B[ 8] ^= Bx[ 8]);
        x09 = (B[ 9] ^= Bx[ 9]);
        x10 = (B[10] ^= Bx[10]);
        x11 = (B[11] ^= Bx[11]);
        x12 = (B[12] ^= Bx[12]);
        x13 = (B[13] ^= Bx[13]);
        x14 = (B[14] ^= Bx[14]);
        x15 = (B[15] ^= Bx[15]);
        for (i = 0; i < 8; i += 2) {
                /* Operate on columns. */
                x04 ^= ROTL(x00 + x12,  7);   x09 ^= ROTL(x05 + x01,  7);
                x14 ^= ROTL(x10 + x06,  7);   x03 ^= ROTL(x15 + x11,  7);

                x08 ^= ROTL(x04 + x00,  9);   x13 ^= ROTL(x09 + x05,  9);
                x02 ^= ROTL(x14 + x10,  9);   x07 ^= ROTL(x03 + x15,  9);

                x12 ^= ROTL(x08 + x04, 13);   x01 ^= ROTL(x13 + x09, 13);
                x06 ^= ROTL(x02 + x14, 13);   x11 ^= ROTL(x07 + x03, 13);

                x00 ^= ROTL(x12 + x08, 18);   x05 ^= ROTL(x01 + x13, 18);
```

```
                x10 ^= ROTL(x06 + x02, 18);  x15 ^= ROTL(x11 + x07, 18);

                /* Operate on rows. */
                x01 ^= ROTL(x00 + x03,  7);  x06 ^= ROTL(x05 + x04,  7);
                x11 ^= ROTL(x10 + x09,  7);  x12 ^= ROTL(x15 + x14,  7);

                x02 ^= ROTL(x01 + x00,  9);  x07 ^= ROTL(x06 + x05,  9);
                x08 ^= ROTL(x11 + x10,  9);  x13 ^= ROTL(x12 + x15,  9);

                x03 ^= ROTL(x02 + x01, 13);  x04 ^= ROTL(x07 + x06, 13);
                x09 ^= ROTL(x08 + x11, 13);  x14 ^= ROTL(x13 + x12, 13);

                x00 ^= ROTL(x03 + x02, 18);  x05 ^= ROTL(x04 + x07, 18);
                x10 ^= ROTL(x09 + x08, 18);  x15 ^= ROTL(x14 + x13, 18);
        }
        B[ 0] += x00;
        B[ 1] += x01;
        B[ 2] += x02;
        B[ 3] += x03;
        B[ 4] += x04;
        B[ 5] += x05;
        B[ 6] += x06;
        B[ 7] += x07;
        B[ 8] += x08;
        B[ 9] += x09;
        B[10] += x10;
        B[11] += x11;
        B[12] += x12;
        B[13] += x13;
        B[14] += x14;
        B[15] += x15;
}
void scrypt_1024_1_1_256_sp_generic(const char *input, char *output, char
*scratchpad)
{
        uint8_t B[128];
        uint32_t X[32];
        uint32_t *V;
        uint32_t i, j, k;

        V = (uint32_t *)(((uintptr_t)(scratchpad) + 63) & ~ (uintptr_t)(63));

        PBKDF2_SHA256((const uint8_t *)input, 80, (const uint8_t *)input, 80,
1, B, 128);

        for (k = 0; k < 32; k++)
                X[k] = le32dec(&B[4 * k]);

        for (i = 0; i < 1024; i++) {
                memcpy(&V[i * 32], X, 128);
                xor_salsa8(&X[0], &X[16]);
                xor_salsa8(&X[16], &X[0]);
        }
        for (i = 0; i < 1024; i++) {
                j = 32 * (X[16] & 1023);
                for (k = 0; k < 32; k++)
```

```c
                        X[k] ^= V[j + k];
                xor_salsa8(&X[0], &X[16]);
                xor_salsa8(&X[16], &X[0]);
        }

        for (k = 0; k < 32; k++)
                le32enc(&B[4 * k], X[k]);

        PBKDF2_SHA256((const uint8_t *)input, 80, B, 128, 1, (uint8_t
*)output, 32);
}

#if defined(USE_SSE2)
// By default, set to generic scrypt function. This will prevent crash in
case when scrypt_detect_sse2() wasn't called
void (*scrypt_1024_1_1_256_sp_detected)(const char *input, char *output, char
*scratchpad) = &scrypt_1024_1_1_256_sp_generic;

void scrypt_detect_sse2()
{
#if defined(USE_SSE2_ALWAYS)
    printf("scrypt: using scrypt-sse2 as built.\n");
#else // USE_SSE2_ALWAYS
    // 32bit x86 Linux or Windows, detect cpuid features
    unsigned int cpuid_edx=0;
#if defined(_MSC_VER)
    // MSVC
    int x86cpuid[4];
    __cpuid(x86cpuid, 1);
    cpuid_edx = (unsigned int)buffer[3];
#else // _MSC_VER
    // Linux or i686-w64-mingw32 (gcc-4.6.3)
    unsigned int eax, ebx, ecx;
    __get_cpuid(1, &eax, &ebx, &ecx, &cpuid_edx);
#endif // _MSC_VER

    if (cpuid_edx & 1<<26)
    {
        scrypt_1024_1_1_256_sp_detected = &scrypt_1024_1_1_256_sp_sse2;
        printf("scrypt: using scrypt-sse2 as detected.\n");
    }
    else
    {
        scrypt_1024_1_1_256_sp_detected = &scrypt_1024_1_1_256_sp_generic;
        printf("scrypt: using scrypt-generic, SSE2 unavailable.\n");
    }
#endif // USE_SSE2_ALWAYS
}
#endif

void scrypt_1024_1_1_256(const char *input, char *output)
{
        char scratchpad[SCRYPT_SCRATCHPAD_SIZE];
    scrypt_1024_1_1_256_sp(input, output, scratchpad);
}
```

# Appendix B – Cacti Output for 128kB SRAM Cache

---------- CACTI version 6.0, Non-uniform Cache Access ----------

Optimal number of banks - 32

Grid organization rows x columns - 4 x 8

Average access latency to a random bank

    (Bank Access time + Avg. Network Delay + Contention Cycles)- 77 cycles

Average dynamic energy/access (nJ) - 1.06637

Network frequency - 5 GHz

Cache dimension (mm x mm) - 8.93355 x 15.768

Router stats:

    Maximum possible network frequency - 5.40707 GHz

    Network frequency - 5 GHz

    No. of Virtual channels - 4

    No. of pipeline stages - 3

    Link bandwidth - 64 (bits)

    No. of buffer entries per virtual channel - 8

    Simple buffer access (read) - 0.000147822 (nJ)

    Simple buffer access (write) - 0.000200725 (nJ)

    Cross bar access energy - 0.00108219 (nJ)

    Arbiter access energy - 0.000115668 (nJ)

Wire stats:

    Wire type - Full swing global wires with 30% delay penalty

    Wire width - 64 nm

Wire spacing - 64 nm

Horizontal link delay - 0.207929 (ns)

Vertical link delay - 0.252422 (ns)

Delay/length - 0.105495 (ns/mm)

Horizontal link energy -dynamic/access 0.000131451 (nJ)

-leakage 1.57664 (nW)

Vertical link energy -dynamic/access 0.000159579 (nJ)

-leakage 1.91402 (nW)

Energy/length per wire - 7.14518e-05 (nJ/mm)

Detailed Bank Stats:

Bank Size (bytes): 2097152

Number of banks: 1

Associativity: 8

Block size (bytes): 128

Read/write Ports: 1

Read ports: 0

Write ports: 0

Technology size: 0.03um

Access time (ns): 0.75198

Cycle time (ns):  0.147279

Total dynamic read energy per access (nJ):0.121163

Total leakage power of a bank (mW):1192.34

Cache height x width (mm): 2.233379 x 1.839713

Best Ndwl (L1): 8

Best Ndbl (L1): 32

Best Nspd (L1): 0.500000

Best Ndcm (L1): 8

Best Ndsam (L1): 4

Best Ntwl (L1): 2

Best Ntbl (L1): 4

Best Ntspd (L1): 2.000000

Best Ntcm (L1): 1

Best Ntsam (L1): 4

Data array, H-tree wire type: Global wires with 20% delay penalty

Tag array, H-tree wire type: Global wires with 30% delay penalty

Time Components:

Data side (with Output driver) (ns): 0.75198

    H-tree input delay (ns): 0.288951

    Decoder + wordline delay (ns): 0.0896385

    Bitline delay (ns): 0.0385681

    Sense Amplifier delay (ns): 0.03

    H-tree output delay (ns): 0.304822

Tag side (with Output driver) (ns): 0.318609

    H-tree input delay (ns): 0.055129

    Decoder + wordline delay (ns): 0.0826787

Bitline delay (ns): 0.0425134

Sense Amplifier delay (ns): 0.03

Comparator delay (ns): 0.0116144

H-tree output delay (ns): 0.0966737

Power Components:

Data array: Total dynamic read energy/access (nJ): 0.104652

Total leakage read/write power all banks at maximum frequency (mW): 1152.16

Total energy in H-tree (that includes both address and data transfer) (nJ): 0.0947923

Decoder (nJ): 0.000141478

Wordline (nJ): 0.000422674

Bitline mux & associated drivers (nJ): 0.000255944

Sense amp mux & associated drivers (nJ): 0.000124171

Bitlines (nJ): 0.00627315

Sense amplifier energy (nJ): 0.00110592

Sub-array output driver (nJ): 0.00128588

Tag array: Total dynamic read energy/access (nJ): 0.0165109

Total leakage read/write power all banks at maximum frequency (mW): 40.1795

Total energy in H-tree (nJ): 0.0111652

Decoder (nJ): 5.22869e-05

Wordline (nJ): 6.28864e-05

Bitline mux & associated drivers (nJ): 3.49353e-05

Sense amp mux & associated drivers (nJ): 0

Bitlines (nJ): 0.00154038

Sense amplifier energy (nJ): 0.00055296

Sub-array output driver (nJ): 0.00291354

Area Components:

Data array: Area (mm2): 3.66352

      Height (mm): 2.23338

      Width (mm): 1.64035

Area efficiency (Memory cell area/Total area) - 68.4658

Tag array: Area (mm2): 0.114996

      Height (mm): 0.576813

      Width (mm): 0.199364

Area efficiency (Memory cell area/Total area) - 68.1618

# Works Cited

Alpha Technologies. (n.d.). *Implementation and Analysis of Scrypt Algorithm in FPGA.* Retrieved from Alpha Technologies: https://alpha-t.net/wp-content/uploads/2013/11/Alpha-Technology-Scrypt-Analysis-on-FPGA-proof-of-concept.pdf

Dillibabu Mannem, S. R. (n.d.). *Performance ANalysis of Reconfigurable SRAM Cell for Low Power Applications.* International Journal of Electronics and Computer Science Engineering.

Nakamoto, S. (2009). Bitcoin: A Peer-to-Peer Electronic Cash System. Retrieved from Bitcoin.

Penn State. (2013). Utah Arch.

Percival, C. (2009). *Stronger Key Derivation Via Sequential Memory-Hard Functions*. Retrieved from tarsnap.com: http://www.tarsnap.com/scrypt.html

Tamai, S. (2013). *Analysis of Bit Cost for STacked Type MRAM.* Yokohama, Japan: Hikari Ltd.

Technology, N. I. (n.d.). *Descriptions of SHA-256, SHA-384, and SHA-512.* Retrieved from Computer Security Division - Computer Security Resource Center: http://csrc.nist.gov/groups/STM/cavp/documents/shs/sha256-384-512.pdf

Wiki, B. (n.d.). *Bitcoin Weaknesses.* Retrieved from Bitcoin Wiki: https://en.bitcoin.it/wiki/Weaknesses#Attacker_has_a_lot_of_computing_power

Wiki, B. (n.d.). *Blocks.* Retrieved from Bitcoin Wiki: https://en.bitcoin.it/wiki/Blocks

Xie, Y. (n.d.). *Emerging NVM Memory Technologies.* Pennsylvania State University.